\documentclass[prc,aps,showkeys,showpacs,nofootinbib,twocolumn]{revtex4}

\usepackage{graphicx}

\usepackage{color} 
\begin{document}
\date{\today}
\title{Projected Quasi-particle Perturbation theory}

\author{Denis Lacroix} \email{lacroix@ganil.fr}  
\affiliation{GANIL, CEA/DSM and CNRS/IN2P3, Bo\^ite Postale 55027, 14076 Caen Cedex, France}  
\author{Danilo Gambacurta}  
\affiliation{GANIL, CEA/DSM and CNRS/IN2P3, Bo\^ite Postale 55027, 14076 Caen Cedex, France}  

\begin{abstract}
The BCS and/or HFB theories are extended by treating the effect of four quasi-particle states perturbatively. 
The approach is tested on the pairing hamiltonian, showing that it combines the advantage of standard 
perturbation theory valid at low pairing strength and of non-perturbative approaches breaking particle number valid at 
higher pairing strength. Including the restoration of particle number, further improves the description of pairing correlation. 
In the presented test, the agreement between the exact solution and the combined perturbative + projection 
is almost perfect. The proposed method scales friendly when the number of particles increases and provides a simple alternative 
to other more complicated approaches. 
\end{abstract}

\date{\today}


\maketitle

\section{Introduction}

Theories breaking explicitly the particle number, like Bardeen-Cooper-Schrieffer
(BCS) or Hartree-Fock-Bogoliubov (HFB) are the simplest efficient ways to describe 
pairing correlations in physical systems. However, these theories suffer
from some limitations especially when the number of particles becomes rather small as 
it is the case in mesoscopic systems like in nuclear physics \cite{Rin80} or condensed matter \cite{Von01}.
The main difficulty is the sharp transition from normal to superfluid phases as the pairing strength 
increases. The second difficulty is a systematic underestimation of pairing correlations that increases 
when the particle number decreases.    

In the last decades, several approaches have been used to overcome these difficulties. When the number 
of particle is small enough, the exact solution of the pairing problem is accessible by direct diagonalization of the 
Hamiltonian \cite{Vol01,Zel03,Vol07} 
and/or using the secular equation originally proposed by Richardson \cite{Ric64,Duk04,Van06}. 
Accurate description of the ground state energy of superfluid systems with larger particle 
number can be obtained using either Quantum Monte-Carlo technique (see for instance \cite{Cap98,Muk11}) 
or extending quasi-particle theories using Variation After Projection approaches \cite{Die64,Hup11} (for recent 
applications see \cite{Rod07,Rod10,Hup12}). All of these methods are however rather involved and demanding 
in terms of computational power. The aim of the present work is to show that a perturbative approach can eventually 
provide a simpler and yet accurate alternative to treat the pairing problem.  

In the following, standard perturbation theory is first recalled. It is then explained how to take advantage of 
both perturbative approaches and BCS/HFB theory. The importance of particle number restoration is finally highlighted.

\section{Standard perturbation theory}

Our starting point here is similar to the one used in refs. \cite{San08,Muk11}.  
We assume a two-body pairing hamiltonian given by: 
\begin{eqnarray}
H & = & \sum_{i=1}^{\Omega} \varepsilon_i (a^\dagger_i a_i + a^\dagger_{\bar i} a_{\bar i} )
+ \sum_{i \neq j}^\Omega v_{ij} a^\dagger_i a^\dagger_{\bar i} a_{\bar j}  a_j  \label{eq:tb} \\
&\equiv& H_0 + V. \nonumber
\end{eqnarray}
Here, $(i,{\bar i})$ denotes time-reversed pair. Such an hamiltonian can be for instance constructed starting from a realistic self-consistent mean-field calculation providing single-particle energies and two-body matrix elements  \cite{Muk11}.

Here $\Omega$ denotes the maximal number of accessible levels for the pairs of particles. In the following, we will 
consider the specific case of equidistant levels with $\varepsilon_i  = (i-1) \Delta \varepsilon $ with
$\Delta \varepsilon = 1$ MeV and a number of particles $N= \Omega$.

As already noted some times ago \cite{Bis71} and recently rediscussed \cite{Sch01, San08},
the two-body part $V$ can be treated perturbatively to provide an accurate description of the pairing 
problem in the weak pairing interaction regime. Starting from the ground state $| \Phi_0 \rangle$ of $H_0$,
that corresponds to the Slater determinant obtained by occupying the lowest single-particle states while  other 
excited states $| \Phi_n \rangle$ of $H_0$ can be obtained by considering particle-hole (p-h), 2p-2h, ... excitations built on top of the 
$| \Phi_0 \rangle$. 
Using second-order perturbation theory, the ground state energy of $H$ reads:
\begin{eqnarray}
{\mathbf E}_0 & = & E_0 + E^{(2)}_0
\end{eqnarray} 
where $E_0 = 2 \sum_{i=1,N_{\rm pair}} \varepsilon_i$ (with $N_{\rm pair}=N/2$) is the ground state 
energy of $H_0$ while $E^{(2)}_0$ is the standard second-order correction, that can be found in textbook:
\begin{eqnarray}
E^{(2)}_0 &=& \sum_{n \neq 0} \frac{\left| \langle \Phi_0 |V| \Phi_n \rangle \right|^2  }{E_0 - E_n}
\label{eq:standardpt}
\end{eqnarray}
In the specific case considered here, i.e. Eq. (\ref{eq:tb}), only 2p-2h states contribute to the second order correction. 
More precisely, the states $n$ could only differ from the ground state 
by one occupied pair  above the Fermi sea (denoted by $j$) and one unoccupied pair below (denoted by $i$):
$| \Phi_{i,j} \rangle  = a^\dagger_j a^\dagger_{\bar j} a_{\bar i} a_{i} | \Phi_{0} \rangle$ and are eigenstates 
of $H_0$ with energy $E_{ij}  =  E_0 + 2 ( \varepsilon_j - \varepsilon_i)$. 
This leads to 
\begin{eqnarray}
E^{(2)}_0 = - \frac{1}{2} \sum_{i=1,N_{\rm pair}} \sum_{j = N_{\rm pair}+1 , N_{\rm max}} \frac{|v_{ij}|^2}{\varepsilon_j - \varepsilon_i} ,\label{eq:corr2}
\end{eqnarray}  
with $N_{\rm pair} = N/2$ and $N_{\rm max}=\Omega/2$. This expression has been obtained in ref. \cite{Sch01}
using a different approach based on the Richardson-Gaudin equation. It is known \cite{Bis71,San08} that standard
perturbation theory provides an appropriate description in the weak coupling regime.  

\begin{figure}[htbp] 
\includegraphics[width=0.9\linewidth]{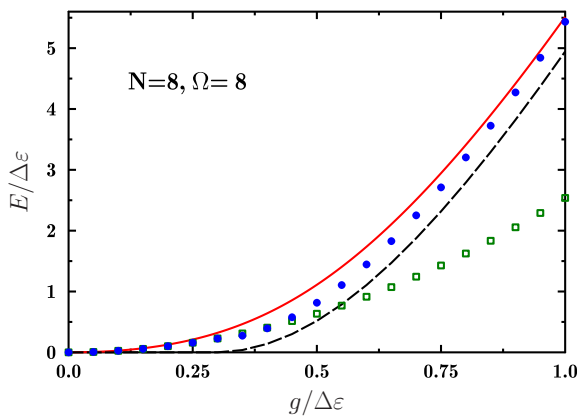} 
\caption{
(color online) Illustration of the correlation energy as
a function of the coupling strength for the 
case of N=8 particles and $\Omega=8$. 
The exact result solution (red solid curve), BCS (black dashed line), 
standard perturbation theory (green open squares), 
Eq. (\ref{eq:standardpt}) and QP$^2$T theory with second order correction  Eq. (\ref{eq:qppt}) (blue filled circles), are displayed.}
\label{fig1:pert} 
\end{figure} 

In figure \ref{fig1:pert}, an example of standard perturbation theory (SPT) is presented for the $N=8$ particles and constant coupling case, i.e.  
$v_{ij} = - g$. The correlation energy defined as the difference between the Hartree-Fock energy $E_0$ and the ground state 
energy ${\mathbf E}_0$ obtained with STP are compared to the exact solution and BCS result.  
In the latter case, pairing correlation is non-zero only above the threshold value $g/\Delta \varepsilon \simeq 0.3$. As illustrated 
from Fig. \ref{fig1:pert}, standard perturbation theory matches with the exact result below the threshold but significantly underestimates 
the correlation for larger $g$ value. This aspect underlines the highly non-perturbative nature of the pairing quantum phase-transition.
On the opposite, one of the advantage a theory like BCS is the possibility to incorporate non-perturbative physics even in the strong interaction case by breaking the U(1) symmetry
associated to particle number conservation.     

\section{Quasi-particle Perturbation theory}

To provide a proper description of both the weak and strong pairing strength regime, it seems quite natural 
to try to combine theories based on quasi-particles and perturbative approaches. This possibility has been 
explored long time ago in Refs. \cite{Bal62,Hen64} mainly to discuss the removal of "dangerous diagram" occuring
in normal perturbation theory. Note that recently, it has also been revisited as a possible tool to perform 
ab-initio calculation in nuclei \cite{Som11} based on Gorkov-Green function formalism \cite{Gor58}.  
In the following, it is assumed that the BCS/HFB approach has been applied in a preliminary study 
and that the hamiltonian (\ref{eq:tb}) is written in the canonical basis of the quasi-particle ground state. Then, the ground 
state takes the form 
\begin{eqnarray}  
|\Phi_0 \rangle = &=&  \prod_{i>0} \left( U_i + V_i a^\dagger_i a^\dagger_{\bar  i}\right)| - \rangle,  \label{eq:phi0standard} 
\end{eqnarray}
and is the vacuum of the quasi-particle creation operators 
defined through: 
\begin{eqnarray}
\beta^\dagger_i & = & U_i a^\dagger_i - V_i a_{\bar i}, \\
\beta^\dagger_{\bar i} & = & U_i a^\dagger_{\bar i}  + V_i a_{i} .
\end{eqnarray}
In the HFB/BCS theory, the original hamiltonian is replaced by an effective Hamiltonian 
that is conveniently written as \footnote{Note that here, it is implicitly assumed that $H$ is replaced by $H - \lambda N$}:  
\begin{eqnarray}
H_0 &=& E_0 + \sum_{i} E_i \left( \beta^\dagger_i \beta_i +   \beta^\dagger_{\bar i} \beta_{\bar i} \right) \label{eq:hqp}
\end{eqnarray}
where $E_0$ is the BCS/HFB ground state energy, while $E_i$ corresponds to the quasi-particle energy given by:
\begin{eqnarray}
E_i  & = & \sqrt{(\varepsilon_i - \lambda )^2 + \Delta^2}
\end{eqnarray}
where $\lambda$ is the Lagrange multiplier used to impose the average particle number while $\Delta$ is the pairing gap 
(for a detailed discussion see \cite{Bri05}).

For even systems, excited states of $H_0$ are 2 quasi-particle (2QP), 4 quasi-particle (4QP), ... excitations with respect to the ground state. 
The original hamiltonian $H$ contains many terms that are neglected in $H_0$ \cite{Rin80} and that are responsible from the deviation 
between the quasi-particle and the exact solution. However, noting that:
\begin{eqnarray}
H | \Phi_{0} \rangle  = \left(H_0 - \sum_{i \neq j} v_{ij} U^2_i V^2_j \beta^\dagger_i \beta^\dagger_{\bar i}   \beta^\dagger_j \beta^\dagger_{\bar j} \right)  | \Phi_{0} \rangle,
\end{eqnarray}   
it can be anticipated that the main source of discrepancy is due to the coupling of $\Phi_0$ with the 4QP states.   Some arguments 
showing that 4QP states should improved the description of pairing especially in the weak coupling regime have been 
given in ref. \cite{Man66}.

Below, perturbation theory is applied assuming that $H_0$ (Eq. \ref{eq:hqp}) is the unperturbed hamiltonian while the perturbation $V$ 
is given by: 
\begin{eqnarray}
V &=& - \sum_{i \neq j} v_{ij} U^2_i V^2_j \beta^\dagger_i \beta^\dagger_{\bar i}   \beta^\dagger_j \beta^\dagger_{\bar j}. 
\end{eqnarray}
$V$ couples the ground state with the 4QP states, defined as (for $i > j$):
  \begin{eqnarray}
|\Phi_{i,j} \rangle & = &  \beta^\dagger_i \beta^\dagger_{\bar i}  \beta^\dagger_j \beta^\dagger_{\bar j}  |\Phi_{0} \rangle
\end{eqnarray}
and associated to the unperturbed energy 
\begin{eqnarray}
E_{ij} = E_0 + 2 \left(E_{i} + E_{j} \right).
\end{eqnarray}
The present approach, that is a direct extension of standard perturbation theory, is 
called hereafter quasi-particle perturbation theory (QP$^2$T). Using Eq. (\ref{eq:standardpt}), the second order 
correction to the ground state energy is equal to:
\begin{eqnarray}
E^{(2)}_0 &=&  - \frac{1}{2} \sum_{i > j} v^2_{ij} \frac{ (U^2_i V^2_j + U^2_j V^2_i )^2}{E_i + E_j}. \label{eq:qppt}
\end{eqnarray} 
This correction properly extends Eq. (\ref{eq:corr2}) from the normal to the superfluid phase. Indeed at the threshold value 
of $g$, i.e. when $\Delta \rightarrow 0$, the 4QP states identify with 2p-2h excitations while:
\begin{eqnarray}
E_i + E_j & \rightarrow & |\varepsilon_i - \lambda | + |\varepsilon_j - \lambda | = \varepsilon_i - \varepsilon_j,
\end{eqnarray}   
and the standard perturbation theory case is recovered. 

The result obtained with the QP$^2$T approach at second order in perturbation (\ref{eq:qppt}) are displayed 
in figure \ref{fig1:pert} with filled circles. Note that below the BCS threshold, standard perturbation theory 
is used. The present approach can be regarded as a rather academic exercise but it turns out to provide a very simple 
way to extend mean-field theory based on quasi-particle states. In particular, it avoids the threshold problem of the latter 
and improves the description of correlation in the intermediate and strong coupling case.  

\section{Effect of the restoration of particle number}

Similarly to the original quasi-particle theory, the energy deduced from the QP$^2$T contains spurious
contribution coming from the fact that the perturbed state does not preserves particle number.   Indeed, 
using standard formulas, to second order in perturbation, the ground state expresses as:
\begin{eqnarray}
| \Psi_0 \rangle &=&   | \Psi^{(0)}_0 \rangle + | \Psi^{(1)}_0 \rangle +  | \Psi^{(2)}_0 \rangle ... \nonumber \\
&=& | \Phi_0 \rangle + \sum_{i>j} c_{ij} | \Phi_{i,j} \rangle + \cdots \label{eq:statepert}
\end{eqnarray} 
where $| \Psi^{(i)}_0 \rangle$, $i=0$, $1$, ... denotes the contribution to the state at $i^{th}$ order in perturbation 
and where 
\begin{eqnarray}
c_{ij} &=&  - \frac{v_{ij}} {2} \frac{ (U^2_i V^2_j + U^2_j V^2_i) }{E_i + E_j} . \label{eq:mix}
\end{eqnarray}
Above the BCS threshold, neither  $| \Phi_0 \rangle$ nor $| \Phi_{i,j} \rangle$ are eigenstates
of the particle number operator. 

The most direct way to remove spurious contributions due to the mixing of different particle number is to introduce the
operator $P^N$
\begin{equation}
\label{eq:Pop}
{P}^N
=  \frac{1}{2\pi} \int_{0}^{2\pi} \! d{\varphi} \; \,e^{i\varphi (\hat{N}-N)}.
\end{equation}  
that projects onto particle number $N$. 

\begin{figure}[htbp] 
\includegraphics[width=0.9\linewidth]{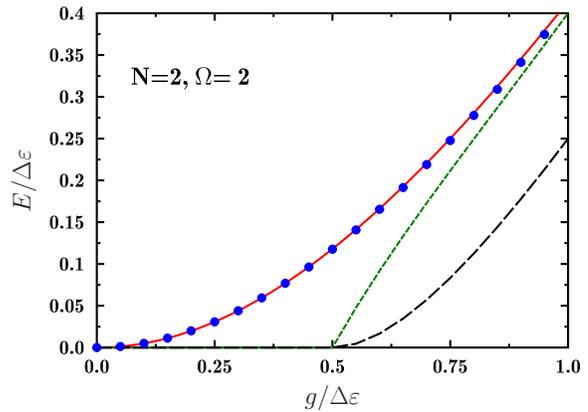} 
\caption{(color online) Illustration of the correlation energy as
a function of the coupling strength for the 
case of N=2 particles and $\Omega=2$. 
The exact result solution (blue solid line), PAV result (green short-dashed line)
and the combination of projection and perturbation theories
(red filled circles) are shown. The original BCS (black long dashed line) is also shown as a reference. }
\label{fig:projN2} 

\end{figure} 

\begin{figure}[htbp] 
\includegraphics[width=0.9\linewidth]{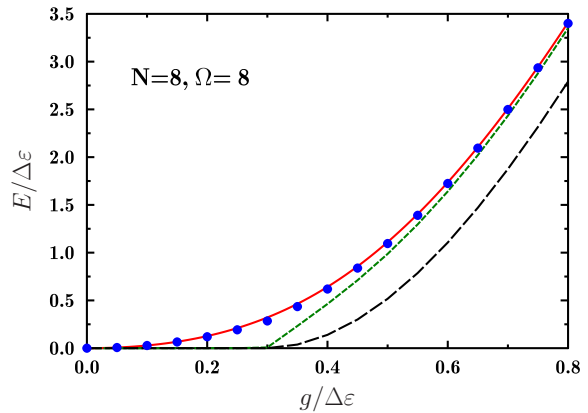} 
\caption{
Same as figure \ref{fig:projN2} for the $N=\Omega=8$ case.
}
\label{fig:projN8} 
\end{figure} 
\begin{figure}[htbp] 
\includegraphics[width=0.9\linewidth]{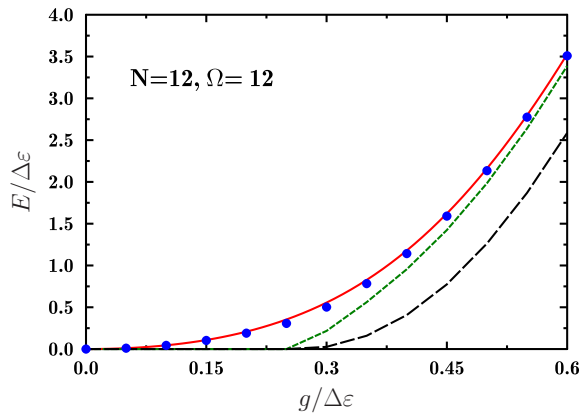} 
\caption{
(color online) Same as figure \ref{fig:projN2} for the $N=\Omega=12$ case.}
\label{fig:projN12} 
\end{figure} 

The most straightforward way to combine projection with quasi-particle perturbation 
theory is to directly take the expectation value 
\begin{eqnarray}
{\rm \mathbf E}_0 & = & \frac{\langle \Psi^N_0 |H| \Psi^N_0\rangle }{\langle  \Psi^N_0 | \Psi^N_0 \rangle }  \label{eq:e0proj}
\end{eqnarray}
with $| \Psi^N_0 \rangle  =  P^N | \Psi_0 \rangle$ and where $| \Psi_0 \rangle$ is truncated at a given order in 
perturbation. This approach will be referred as the projected quasi-particle perturbation
theory (QP$^3$T) in the following. When only the zero order in perturbation 
is retained in Eq. (\ref{eq:statepert}), the QP$^3$T identifies with the projection after variation (PAV) that is commonly used, especially in the nuclear Energy Density Functional approach (EDF) \cite{Ben03}.
Formulas useful to compute the expectation values of one- and two-body operators with projection are given 
in appendix \ref{app:proj}. 
The projection
is performed numerically using these expressions and
the Fomenko discretization procedure of the gauge-space
integrals \cite{Fom70,Ben09}. Here, $199$
discretization points have been used. Note that the number of points 
can be reduced down to $5$ without changing the result.   
The correlation energies obtained for $N=2$, $N=8$ and $N=12$ (each time with $\Omega=N$) using the second order QP$^3$T  
are shown in figures \ref{fig:projN2}, \ref{fig:projN8} and \ref{fig:projN12} respectively. In each case, the original BCS, the PAV
and the exact solution are also shown. 
The result of QP$^3$T almost superimposed with the exact solution. Only a slight difference can be seen 
around the BCS threshold.  

\begin{figure}[htbp] 
\includegraphics[width=0.9\linewidth]{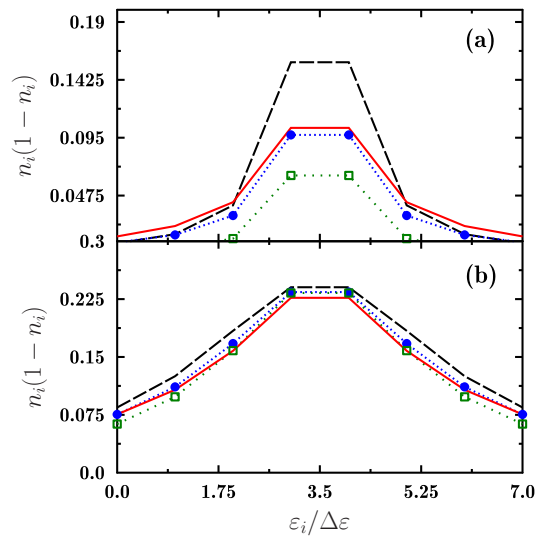} 
\caption{(color online) Example of evolution of the quantity $n_i (1-n_i)$ obtained with QP$^3$T (filled blue circles)
as a function of single-particle energy 
for the $N=\Omega=8$ case with $g/\Delta \varepsilon= 0.4$ (a) and $0.8$ (b). In both case, the exact result (red solid line) 
the BCS result (black dashed line) and the PAV result (green open squares) are shown. }
\label{fig:occN8} 
\end{figure}

Besides the energy, other observables can also be estimated. As an example, the quantity 
$n_i (1-n_i)$ is shown in figure \ref{fig:occN8} as a function of single-particle energies, where 
$n_i$ are the occupation numbers of single-particle states. This quantity is a measure of the deviation from the independent
particle picture where it is strictly zero for all states. The use of perturbation and projection considerably 
improves the description of one-body observables especially at intermediate coupling where the original BCS deviates significantly from
the exact solution. In addition, even if the PAV differs significantly from the exact case as it was noted in ref. \cite{Hup11}, 
the extra mixing with the 4QP states compensates this drawback of PAV.    

In view of this agreement, it seems that the QP$^3$T does automatically select important many-body states, namely projected ground state 
and projected 4QP states, on which the true eigenstate decomposes.  These states are highly non-trivial multi-particle multi-hole 
mixing that can also be describe by direct diagonalization of the Hamiltonian but that is much more demanding numerically. 
Indeed, the size of the 4QP Hilbert space is $\Omega (\Omega - 1)/2$ while the size of the matrix to diagonalize the hamiltonian 
is  ($\Omega !/[N_{\rm pair}!(\Omega - N_{\rm pair})!]$). 
It is important to recall that here no diagonalization is required 
since the mixing coefficients are directly given by the quasi-particle perturbation theory, Eq.  (\ref{eq:mix}).  These features make the 
approach rather simple to implement on existing HFB/BCS codes to provide a much better approximation than the PAV that is often 
currently used. In addition, by contrast  to the Variation After Projection that is rather involved \cite{Rod07,Hup12}, no extra 
minimization is required. 
 
\begin{figure}[htbp] 
\includegraphics[width=0.9\linewidth]{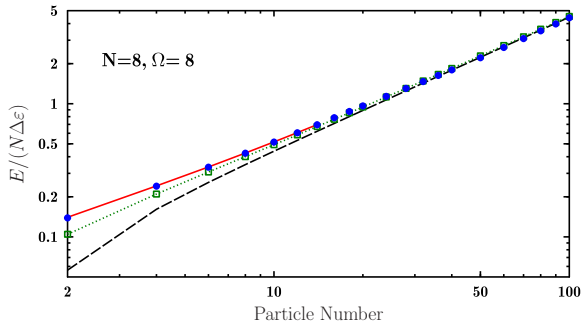} 
\caption{(color online)  Evolution of the correlation energy per particle as a function of the number of particles for
$g/\Delta \varepsilon = 0.8$
for the BCS (black dashed line), QP$^2$T (green open squares) and QP$^3$T (blue filled circles). 
The exact case (solid red line) is also shown up to $N=14$ particles. }
\label{fig:systN} 
\end{figure}

Contrary to the exact diagonalization, the  QP$^2$T and QP$^3$T can be performed even for large particle number. 
As an illustration, in figure \ref{fig:systN}, a systematic study of the correlation energy evolution obtained with some of the 
approaches presented above as the number of particle increases up to $N=100$ for the case $g/\Delta \varepsilon = 0.8$. With standard diagonalization techniques, the exact solution can hardly be obtained for $N>14$.  Other approaches based on quasi-particle 
theories can be applied without difficulties. 
Note however, that the QP$^3$T requires to perform more and more gauge angles integrations  (see appendix \ref{app:proj})
as $N$ increases making the calculations more time consuming with respect to the non-projected theories like BCS or QP$^2$T.

From this figure, we also note that differences between theories are seen only for rather small particle number $N<30$, while above 
BCS and  QP$^3$T cannot be distinguished.  However, as it has been discussed previously, for $N<30$, the  QP$^3$T is the 
only theory that can provide an excellent reproduction of the exact result when available.

\section{Summary}

In this work an extension of the standard many-body theory to treat the pairing 
problem is introduced. Including the effect of 4QP pertubatively to extend usual BCS/HFB removes 
the problem of sharp transition from normal to superfluid phase and significantly 
improves the description of pairing. Last, when restoration of particle number is performed within the 
perturbative approach, a perfect agreement with the exact results is found.  This finding provide 
a direct proof of the importance of 4QP state to extend mean-field theories.
In addition, it is shown that the 
quasi-particle perturbation theory can be implemented even for large particle number without special difficulties, making 
the technique rather attractive and much simpler than other approaches, like full diagonalization, variation after projection or 
Quantum Monte-Carlo techniques. Last, it is worth mentioning that the present technique can be directly and rather easily implemented on existing BCS/HFB codes  \cite{Bon05,Doba04} to improve the description of pairing correlations.  

Recently, the use of Gorkov-Green function theory has been proposed \cite{Som11} as a possible 
tool to perform ab-initio calculation for nuclei. This theory provides a general formalism based 
on quasi-particle states. The result obtained in the present study are rather encouraging to pursue
in that direction and that projection might be needed. 

\section*{Acknowledgment}  
We thank G. Bertsch for helpful discussion and for providing the code to perform the exact diagonalization of
the pairing problem. We also thank G. Hupin for proofreading the manuscript and T. Duguet for pointing out some 
relevant references. 
 
\appendix 

\section{Expression of projected quantities}
\label{app:proj}

Starting from the standard expression of the quasi-particle ground state (Eq. (\ref{eq:phi0standard})), the 4 QP states
are given by:
\begin{eqnarray}
| \Phi_{i,j} \rangle &=& \left(-V_i + U_i a^\dagger_i a^\dagger_{\bar  i} \right)
\left(-V_j + U_j a^\dagger_j a^\dagger_{\bar  j} \right) \nonumber \\
&& \prod_{k>0, k\neq (i,j)} \left( U_k + V_k a^\dagger_k a^\dagger_{\bar  k}\right)| - \rangle.  
\end{eqnarray}
For compactness, this expression is written as 
\begin{eqnarray}
| \Phi_m \rangle &=& \prod_{k>0} \left( U^m_k + V^m_k a^\dagger_k a^\dagger_{\bar  k}\right)| - \rangle.  
\end{eqnarray}
This notation includes the ground state case ($m=0$). The state obtained in QP$^3$T can be generically written as 
$| \Psi^N_0 \rangle = \sum_m c_m  P_N| \Phi_m \rangle$, and for any operator $O$ that conserves the particle number, we have 
\begin{eqnarray}
\langle \Psi^N_0 |O | \Psi^N_0 \rangle = \sum_{m,n} c^*_n c_m  \langle \Phi_n |O P^N| \Phi_m \rangle
\end{eqnarray} 
where

\begin{equation}
\label{eq:expproj}
\langle \Phi_n |O P^N| \Phi_m \rangle =  \int_{0}^{2\pi}  d \varphi \, 
\frac{e^{-i\varphi N}}{2\pi} \langle \Phi _n | O  | \Phi_m (\varphi) \rangle,
\end{equation} 
and
\begin{eqnarray}
| \Phi_m (\varphi) \rangle & = &  \prod_{k>0} \left( U^m_k + V^m_k e^{2i\varphi} a^\dagger_k a^\dagger_{\bar  k}\right)| - \rangle.   
\end{eqnarray}
\begin{widetext}
Starting from this expression it could be deduced that:
 \begin{eqnarray}
 \langle \Phi_n | P^N| \Phi_m \rangle & = &  
 \frac{1}{2\pi} \int_{0}^{2\pi} \! d{\varphi} e^{-i\varphi N}  \prod_{k > 0}  \big( U^n_k U^m_k + V^n_{k} V^m_{k} e^{2i\varphi} \big) 
\end{eqnarray}
 \begin{eqnarray}
 \langle \Phi_n | a^\dagger_i a_i P^N| \Phi_m \rangle & = &  
 \frac{1}{2\pi} \int_{0}^{2\pi} \! d{\varphi} e^{-i\varphi N}  V^n_{i} V^m_{i} e^{2i\varphi}  
 \prod_{k > 0, k \neq i}  \big( U^n_k U^m_k + V^n_{k} V^m_{k} e^{2i\varphi} \big) \nonumber 
\end{eqnarray}
 \begin{eqnarray}
 \langle \Phi_n | a^\dagger_i a^\dagger_{\bar i} a_{\bar j}  a_j P^N| \Phi_m \rangle & = &  
 \frac{1}{2\pi} \int_{0}^{2\pi} \! d{\varphi} e^{-i\varphi N}  V^n_{i} U^m_i U^n_j  V^m_{j} e^{2i\varphi}  
 \prod_{k > 0, k \neq (i,j)}  \big( U^n_k U^m_k + V^n_{k} V^m_{k} e^{2i\varphi} \big) \nonumber 
 \end{eqnarray} 
 where the latter expression is valid for $i \neq j$. 
 \end{widetext}


\end{document}